# Single-step method for the immobilization of hydroxyapatite on 3D-printed porous polyetherketoneketone implants

*Semen Goreninskii [1,2]\*, Igor Akimchenko [3], Alexander Vorobyev[4], Mikhail Konoplyannikov[5,6], Yuri Efremov[7], Evgeniy Sudarev[8], Andrei Zvyagin[1,6], Evgeny Bolbasov[2,3], Sergei Tverdokhlebov[3]*

1) Onconanotheranostics Laboratory, Shemyakin-Ovchinnikov Institute of Bioorganic Chemistry RAS, 117997 Moscow, Miklukho-Maklaya, st., 16/10, Russia
2) Additive Technologies Center, Tomsk Polytechnic University, 634050 Tomsk, Lenina av., 30, Russia
3) B.P. Weinberg Research Center, Tomsk Polytechnic University, 634050 Tomsk, Lenina av., 30, Russia
4) Research School of Chemistry & Applied Biomedical Sciences, Tomsk Polytechnic University, 634050 Tomsk, Lenina av., 30, Russia
5) Federal Research Clinical Center of Specialized Medical Care and Medical Technologies, Federal Medical-Biological Agency of the Russian Federation, 123182 Moscow, Volokolamsk Highway, 30, Russia
6) Institute of Molecular Theranostics, Sechenov First Moscow State Medical University, 119991 Moscow, Trubetskaya st., 8, Russia
7) Institute for Regenerative Medicine, Sechenov First Moscow State Medical University, 119991 Moscow, Trubetskaya st., 8, Russia
8) N.M. Kizhner Research Center, Tomsk Polytechnic University, 634050 Tomsk, Lenina av., 30, Russia

\* Corresponding author: Semen I. Goreninskii (*sig1@tpu.ru*)

# Table of Contents figure

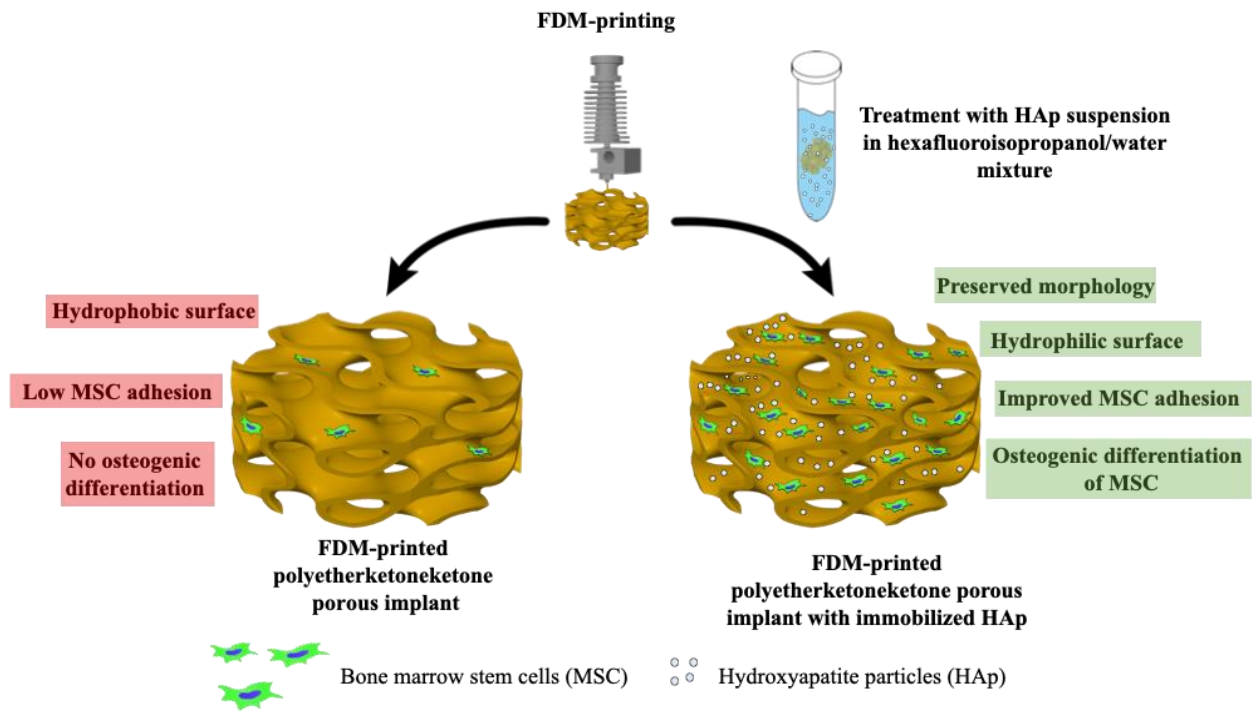

## Abstract

The development of tissue engineering structures (scaffolds) for the reconstruction of bone tissue defects is the relevant task of modern biomedical materials science. Compared to metal-based structures, polymer constructs provide numerous advantages, among them – better processibility and metallosis avoidance. Owing to its high mechanical performance and biocompatibility, polyetherketoneketone (PEKK) became a promising material for the development of such structures. Previously, a method for the immobilization of hydroxyapatite (HAp) on PEKK surface was proposed by our group for the enhancement of stem cell adhesion. In the present study, we propose a single-step method of HAp immobilization on the surface of 3D-printed porous PEKK implants. The proposed approach allowed to preserve the morphology (pore diameter, width of the printed lines) of the pristine implants. With that, up to 35.0±14.0 % of the sample surface were coated with HAp particles, which resulted in improved hydrophilicity (0° water contact angle). The calcium and phosphorus content on the surface of the modified samples was up to 17.4±4.1 and 8.0±1.7 wt. %, respectively. Importantly, the proposed modification preserved compressive strength of the 3D-printed porous PEKK implants. HAp immobilization provided better adhesion of stem cells (from 121±40 cells/mm$^2$ to 234±8 cells/mm$^2$) and induce their osteogenic differentiation.

**Keywords:** hydroxyapatite; polyetherketoneketone; surface modification; bone implant

# 1. Introduction

Tissue engineering combines the approaches of cell biology, medicine and materials science for the regeneration and replacement of the injured organs and tissues. The structures for cells proliferation and adhesion (scaffolds) play key role in tissue engineering [1]. In case of bone tissue restoration, there are specific requirements to scaffolds, especially – high strength and shape defined by defect [2]. For that reason, metals (particularly, titanium and its alloys) became convenient materials for the development of bone tissue implants and scaffolds [3]. However, metals also have numerous disadvantages, such as processing complexity and possibility of metallosis [2,4]. For these reasons, polymer-based implants and scaffolds are being actively developed [5]. Polymer-based implants with needed characteristics (shape and architecture) may be fabricated using FDM-printing technique, they are lightweight and highly biocompatible.

Owing to high mechanical performance and biocompatibility, poly(aryletherketone)s (PAEKs) are promising materials for the fabrication of orthopedic implants [6,7]. Poly(etheretherketone) (PEEK) and poly(etherketoneketone) (PEKK) are the most commonly used polymers of that family. With that, PEKK demonstrates higher biocompatibility [8] and processibility [9]. *In vivo* experiments demonstrate the possibility of PEKK implants application in combination with stem cells for the restoration of bone defects [10–13]. However, the biological inertness of PEKK-based implants results in low osseointegration. For that reason, the methods for modification of PEKK-based implants are being developed [14].

Hydroxyapatite (HAp) is an inorganic component of natural bone tissue. The materials (metals and polymers) coated or doped with HAp are widely applied in orthopedics providing enhanced integration with bone tissue [15,16]. Manzoor et al. extruded a PEKK filament containing 10 wt. % of HAp (including strontium- and zinc-substituted) [17]. Slight drop of the tensile strength was observed for the FDM-printed HAp-doped samples. Converse et al. prepared composite PEKK-based scaffolds containing up to 40 vol. % of HAp using compression molding [18]. The HAp content demonstrated inhomogeneous effect on mechanical properties of the material. The increase of HAp content up to 20 vol. % resulted in the increased Young modulus, while the material containing 40 vol. % lead to its decrease.

Surface modification is a perspective approach for the preservation of mechanical properties of polymer-based implants. Jani et al. used radio-frequency magnetron sputtering for the deposition of strontium-doped coating on PEKK surfaces [19]. Apatite crystals have been grown by Yuan et al. on the surface of porous PEKK samples by pre-etching in sulfuric acid followed by the immersion in simulated body fluid for 5 days [8]. From the push-out test results, the enhanced osseointegration was observed for the samples coated with apatite.

Recently, our group proposed a method for the immobilization of HAp on PEKK surfaces based on the preliminary treatment of the polymer in 1,1,1,3,3,3-hexafluoropropan-2-ol (HFP) [20]. Compared to the convenient methods of HAp deposition, the proposed approach does not require high-cost complex equipment (e.g. magnetrons) and allows the immobilization of HAp in a few minutes. However, the proposed approach includes two steps (sequential stages of polymer pre-treatment and immersion into HAp suspension), which may limit its applications. Moreover, the implants for bone tissue restoration are often porous [21], and that structure should be preserved during the modification process. Thus, the aim of the present study was to develop a single-step method of HAp immobilization on the surface of porous PEKK implants, as well as the investigation of physical, chemical and biological properties of the modified materials.

## 2. Materials and methods

### 2.1 Materials

PEKK filament with a diameter of 1.75 mm (T/I radio 8/2, copolymer 8002PL (Kepstan®, Arkema, France)) was used for the FDM printing of the samples. Chemically pure HFP (99%, P&M Invest, Russia), MilliQ water and HAp powder (Fluidinova, Portugal, 10 μm average particle diameter) were used for the samples modification. For the *in vitro* experiments, alphaMEM medium was purchased from Elabscience (USA), human platelet lysate was purchased from Stemcell Technologies (USA), antibiotic-antimycotic was purchased from Thermo Fischer Scientific (USA), Calcein AM and Hoechst 33342 dyes were purchased from Sigma-Aldrich (USA), MTT reagent was purchased from Servicebio, Ltd. (China), Alizarin Red solution was purchased from Sigma (USA).

### 2.2 Samples fabrication

The samples (discs with 8 mm diameter, 1 mm height and 60 % infill) were fabricated using FDM-printing technique on PEEK-300 3D-printer (CreatBot, China). The following printing regimes were utilized: nozzle diameter of 0.4 mm, nozzle temperature of 390 °C, bed temperature of 150 °C, chamber temperature of 90 °C, printing speed of 30 mm/s, infill pattern – gyroid. For the compression test, tubular samples with 12.7 mm diameter, 25.4 mm height and 60 % infill (according to ISO 604:2002 "Plastics — Determination of compressive properties").

### 2.3 Samples modification

Immobilization of HAp on the surface of PEKK samples was performed in the suspensions of HAp (5, 10 or 20 wt. %) (samples designated as 5%, 10% and 20%, respectively) in HFP/water mixtures (85/15 mass ratio). For the discs modification, 0.5 ml of suspension were poured into 2 ml tubes (Eppendorf, Germany). After that, the PEKK sample was immersed into the suspension for 3 minutes under vortexing. Cylindrical samples prepared for the compression test were modified in 50 ml tubes. The samples were immersed in 10 ml of the suspension for 3 minutes under vortexing. The samples were fully immersed into the suspension during the modification process. After the HAp immobilization, the samples were withdrawn, dryed with filter paper and washed with MilliQ water. The washed samples were dried in VD 115 vacuum furnace (Binder, Germany) under a temperature of 120 °C and 0.08 Pa pressure for 24 h in order to remove the residual HFP.

### 2.4 Morphological studies and investigation of elemental composition

The morphology of the prepared samples was investigated using scanning electron microscopy (SEM) on Vega 3 SBH microscope (Tescan, Czech Republic) equipped with AztecLive Lite Xplore 30 (Oxford Instruments, UK) system for energy dispersive analysis. For the preparation of the samples cross-sections, the samples were frozen in liquid nitrogen and cracked. Before the microscopy, the samples were coated with thin layer of gold using Smart Coater sputtering system (Jeol, Japan). The microscopy was performed in high vacuum at an applied voltage of 10 kV using the secondary electron detector. The morphological characteristics of the samples (pore diameter, width of the printed lines) were measured using ImageJ software (National Institutes of Health, USA). The elemental composition of the prepared samples was investigated using energy dispersive spectroscopy (EDS) in three points of the sample. Samples roughness ($R_a$) was measured using MNP-1 interferential profilometer (TDI SIE SB RAS, Russia). The area of the measurement was 1780×1340 μm, scanning step – 0.5 μm, scanning range – 400 μm. The measurement was performed in triplicates.

### 2.5 Water contact angle measurement

Water contact angle was measured using EasyDrop optical goniometer (Kruss, Germany). To do that, a 3 μl drop of MilliQ water was placed on the surface of the sample. Water contact angle was measured immediately after the placement of the drop. The measurement was performed for five times for each sample.

## 2.6 Compression testing

The compression testing was performed according to ISO 604:2002 "Plastics — Determination of compressive properties" using Instron–1185 (Instron, Norwood, MA, USA) testing machine. Compression speed was set at 1 mm/min. Three samples were studied in each group.

## 2.7 Cell adhesion and differentiation

Bone marrow multipotent stem cells (MSC) (isolated using the standard protocol, approved by the local ethics committee of the FNKTS FMBA of Russia, protocol #11 from 26$^{th}$ of October, 2021) were used for the *in vitro* experiments. Before use, the samples were sterilized in an autoclave. Cell viability was studied as follows. The samples were incubated in MSC culture medium supplemented with 4% of human platelet lysate and 1% of antibiotic-antimycotic for 48 h, then the medium was transferred into 96-well plate with MSC pre-cultured for 48 h. The cells were incubated in the medium for 48 h. Then, MTT reagent was added for 3.5 h, the supernatant was removed and 100 ml of DMSO was added in each well to dissolve the formazan crystals during 15 min. Optical density of the obtained solutions was measured at 580 nm using Victor spectrophotometer (Perkin Elmer, USA). Optical density of the solution obtained from the suspension of cells incubated in pristine medium was taken as control.

For the adhesion studies, the non-modified and modified PEKK samples were placed in the wells of 48-well cell culture plate (Corning, USA) and pre-soaked in the MSC culture medium for 72 h at 37 °C. The cells were seeded from 50 µL of the cell suspension ($30 \times 10^3$ cells) placed on top of each sample. After a 30-min incubation for cell attachment, 350 µL of culture medium were added. MSCs were cultured for 5 days at 37 °C in an atmosphere of 5% $CO_2$, then the proliferation of cells was analyzed. Briefly, the samples were washed with PBS for 5 min and incubated with Calcein AM, and Hoechst 33342 for 30 min at 37°C in the dark, then washed with the culture medium. Fluorescent images were acquired using an EVOS M5000 Imaging System (Thermo Fisher Scientific, Waltham, MA, USA).

The following protocol was used for the investigation of osteogenic differentiation of the MSC. $80 \times 10^3$ cells were cultivated with each sample in wells of 24-well plate for 14 days, the cultivation medium was changed every 3 days. After that, the cells were fixed with 10% formaldehyde solution for 30 minutes. Next, the cells were dyed with Alizarin Red solution for 1 hour. The cells were washed with PBS 3 times.

The images of the cells were captured using PrimoVert inverted microscope with color camera (Carl Zeiss, Germany).

### 2.8 Statistics

Statistical analysis was performed using Prizm 9 software (GraphPad, USA). For the results of morphological, elemental composition and water contact angle studies, Kruskal-Wallis test was used. For the results of the *in vitro* experiment, non-parametric ANOVA test was applied.

### 3. Results and discussion

Morphology of the tissue engineering scaffolds is among the key factors determining their functional characteristics [22]. In our study, the diameter of pores in the model implants was in the range of 400-600 µm, which provides the best cell adhesion and infiltration [13]. SEM images of the pristine and modified PEKK implants are presented in Figure 1.

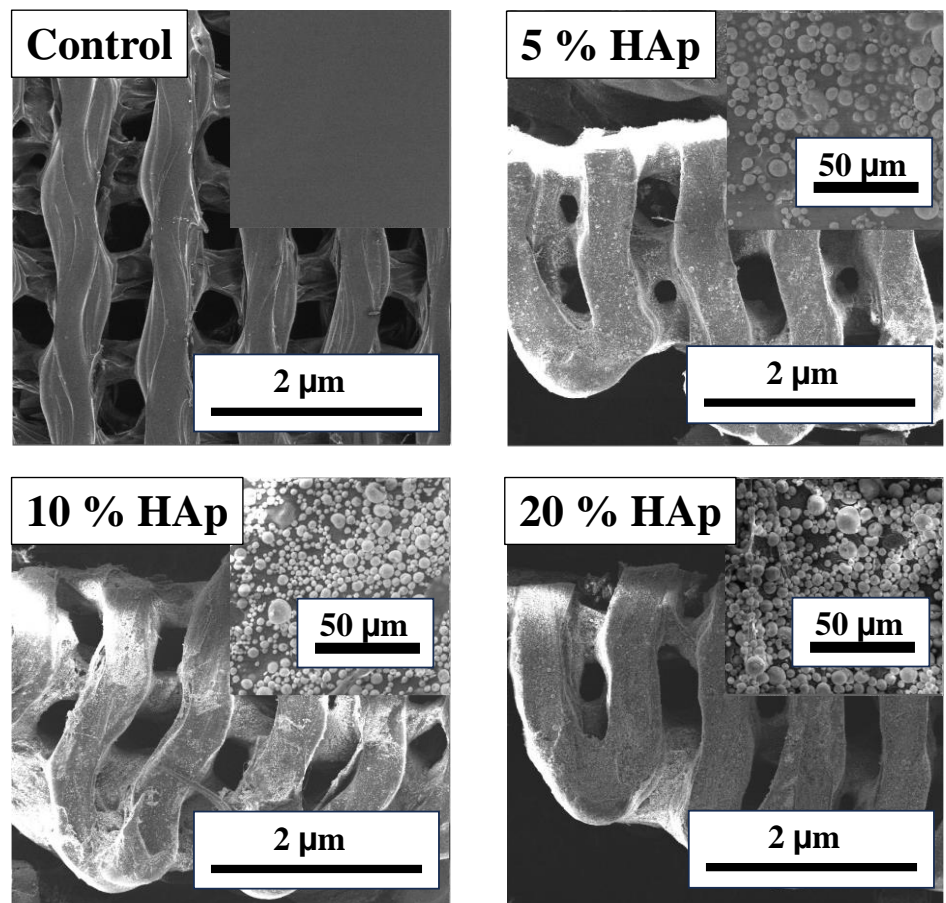

**Figure 1.** SEM images of the samples (magnification ×50, magnification ×2000 in the inserts)

**Table 1.** Morphological parameters of the fabricated samples

| Sample | Pore diameter, μm | Thickness of printed line, μm | HAp-coated area of the sample, % | Roughness ($R_a$), μm |
|---|---|---|---|---|
| Control | 476±24 | 0.42±0.02 | - | 5.0±0.7 |
| 5% HAp | 474±17 | 0.42±0.01 | 10.7±1.0 | 7.5±1.2* |
| 10% HAp | 459±20 | 0.46±0.02 | 27.1±6.8' | 9.3±1.5* |
| 20% HAp | 465±17 | 0.45±0.03 | 35.0±14.0' | 13.4±1.3*' |

' – statistically significant compared to 5% group (Kruskal-Wallis test, p<0.05)
* – statistically significant compared to control group (Kruskal-Wallis test, p<0.05)

The control sample was formed by printed lines with 0.42±0.02 μm thickness and pores with a diameter of 476±24 μm. Despite of the swelling of the polymer surface during the modification process, the thickness of the printed lines and diameter of the pores were not changed significantly (Table 1). With the increase of the HAp content in the modifying suspension, the area coated with HAp increased. The modification in 5 wt. % suspension resulted in coverage of ≈10 % of the sample surface (Table 1). Further increase of the HAp content up to 20 wt. % allowed to cover 35.0±14.0 % of the sample surface. With that, HAp particles did not penetrate the sample bulk and were concentrated on its surface (Figure S1).

Surface roughness is another factor influencing cell adhesion to the implant surface [23]. Roughness of the printed lines forming the control sample was found at 5.0±0.7 μm. The immobilization of HAp particles resulted in the decrease of the roughness up to 13.4±1.3 μm (Table 1), what is comparable to the size of the utilized particles. The presence of HAp on the surface of the modified samples was confirmed by EDS results. The elemental composition of the control sample was presented by carbon and oxygen, the elements forming PEKK macromolecule [24]. In the elemental composition on the sample modified in 5 wt. % suspension, calcium and phosphorus were found (0.6±0.1 and 0.3±0.1 wt. %, respectively) as well as increase of the oxygen content around 1 wt. %. HAp is composed of the listed elements. With the increase of HAp content in the modifying mixture up to 20 wt. %, the concentrations of calcium, phosphorus and oxygen were found at 17.4±4.1, 8.0±1.7 and 39.1±2.8 wt. %, respectively. Ca/P ratio remained constant (around 2.1) for all samples (Table 2). Uneven distribution of the immobilized HAp particles may be explained as follows. It is known that PEKK is a semi-crystalline polymer, what means that both crystal domains and amorphous structures present in the printed samples [25]. It is also known that PEKK crystallinity determines its interactions with water and other solvents [26]. Thus, the inhomogeneous distribution of HAp particles immobilized on the surface of PEKK-based samples may be due to the uneven swelling of their surfaces.

**Table 2.** Elemental composition and water contact angle of the samples

| Sample | Elemental composition, wt. % | | | | | Water contact angle, ° |
|---|---|---|---|---|---|---|
| | C | O | Ca | P | Ca/P | |
| Control | 84.7±0.2 | 15.3±0.2 | - | - | - | 102±3 |
| 5% HAp | 61.8±10.9* | 28.0±6.9* | 6.9±2.5 | 3.4±1.3 | 2.2±0.2 | 0* |
| 10% HAp | 60.0±4.3* | 27.6±4.5* | 8.1±0.2' | 3.7±0.2 | 2.0±0.1 | 0* |
| 20% HAp | 35.5±7.9* | 39.1±2.8* | 17.4±4.1' | 8.0±1.7' | 2.1±0.1 | 0* |

\* – statistically significant compared to control group (Kruskal-Wallis test, $p<0.05$)
' – statistically significant compared to 5% group (Kruskal-Wallis test, $p<0.05$)

Surface hydrophilicity of the implant also influences cell adhesion [27]. The surface of the control sample was found to be hydrophobic (water contact angle of 102±3°), which is typical for porous PEKK materials [28]. The immobilization of HAp in 5 wt. % suspension resulted in hydrophilization of the implant surface (water contact angle of 0°, Table 2). Hydrophilicity is typical for HAp-based coatings [29] and further increase of HAp content in the modifying suspension had no effect on the sample hydrophilicity. Interestingly, in our previous work, where the PEKK surface was pre-treated in HFP vapor, further HAp immobilization did not lead to the surface hydrophilization. We suppose that the pre-treatment in HFP vapor results in better swelling of PEKK surface and the immobilized HAp was partially coated by thin polymer layer.

The control PEKK samples demonstrated extremely low Young modulus and yield strength (5.4±0.4 and 30±2 MPa, respectively) (Table 1), which is significantly lower compared to these parameters for bones [14].

**Table 3.** Compressive properties of the samples

| Sample | Young modulus, MPa | Yield strength, MPa |
|---|---|---|
| Control | 5.4±0.4 | 30±2 |
| 5% HAp | 5.2±0.3 | 31±2 |
| 10% HAp | 5.6±0.5 | 29±2 |
| 20% HAp | 5.2±0.4 | 28±3 |

* – statistically significant compared to control group (Kruskal-Wallis test, $p<0.05$)

Moreover, strength of the fabricated samples was lower compared to porous scaffolds based on poly(lactic acid) [30,31], acrylonitrile butadiene styrene [32], and poly(ε-caprolactone) [33]. We suppose that it is due to the non-optimal architecture (infill and its pattern) of the prepared samples. It is possible to enhance mechanical strength of polymer implants by the optimized architecture providing better load distribution [34–36], or post-treatment [37]. On the other hand, HAp immobilization using the supposed method had no effect on compressive properties of the prepared PEKK-samples (Table 3, Figure S2). Thus, the proposed method allows to preserve mechanical strength of PEKK-based implants.

Since the solvent used for the HAp deposition is toxic, the cytotoxicity of the prepared samples was evaluated (Table 4). The control samples did not demonstrate cytotoxic activity, what is typical for PEKK [14]. With the deposition of HAp particles, a slight decrease of cell viability (≈10-15 %) was observed, which may be due to the trace amount of residual HFP. However, more promising trend was observed in cell adhesion experiment. The results of cell adhesion experiment are presented in Table 4 and Figure 2 (top row).

**Table 4.** Number of adhered cells and viability of cells incubated with the prepared samples

| Sample | Cell viability, % | Number of adhered cells, cells/mm$^2$ |
|---|---|---|
| Control | 100±1 | 121±40 |
| 5% HAp | 94±3 | 208±41* |
| 10% HAp | 84±2* | 220±22* |
| 20% HAp | 90±3 | 234±8* |

* – statistically significant compared to control group (non-parametric ANOVA test, $p<0.05$)

Numerous cells adhered to the surface of the control sample were observed (cell density was found at 121±40 cells/mm$^2$, Figure 2 (top row)). The elongated cells with multiple pseudopodia actively interacted with each other. The deposition of HAp particles from 5 % suspension resulted in 1.7-fold increase of the number of the adhered cells (cell density was found at 208±41 cells/mm$^2$). The same elongated shape of the cells was observed. Further increase of the HAp particles concentration in the modifying mixture did not result in statistically significant changes of the number of the adhered cells (Table 4).

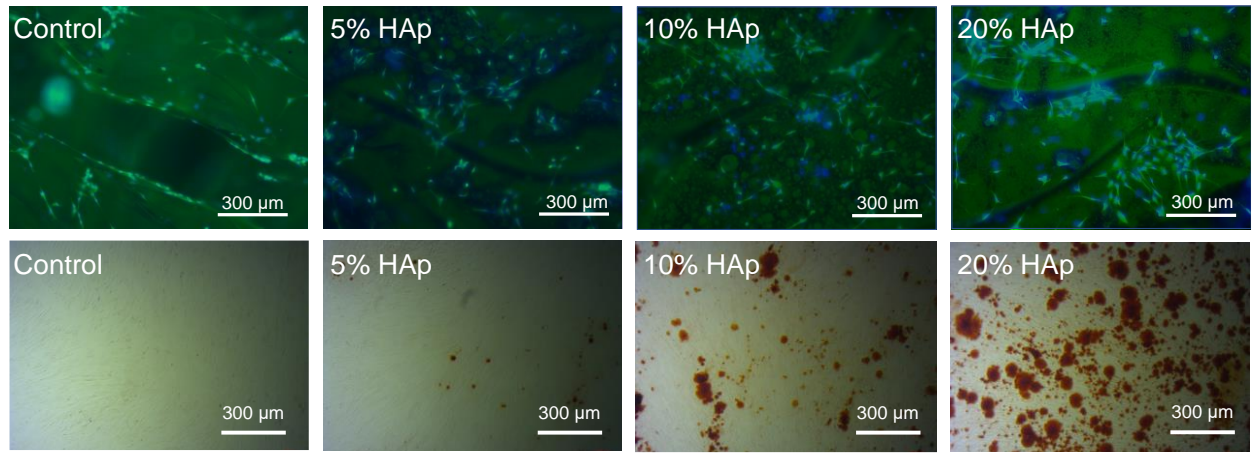

**Figure 2.** Fluorescent microscopy of the cells adhered to the surface of the samples (top row) and optical micrographs of the cells dyed with Alizarin Red after the incubation with the samples (bottom row) (magnification of ×40)

Thus, the deposition of HAp particles using the proposed single-step method, improved the adhesion of stem cells to the surface of the FDM-printed porous PEKK-based implants. The observed effect may be a result of multiple factors. First, it is known, that calcium phosphate-based coatings improve the adhesion and proliferation of stem cells [38]. Second, it was demonstrated that the immobilization of HAp particles using the proposed strategy, increases the roughness of the PEKK-based implants (Table 1). It is also known, that the increased roughness enhances the cell adhesion [39]. Finally, compared to the control PEKK implants, the samples with immobilized HAp particles demonstrated highly hydrophilic surface, which also enhances the adhesion of cells [40]. Thus, using the proposed modification route, it is possible to create optimal conditions for the improved cell adhesion to the surface of PEKK-based implants. With demonstrated possibility to preserve the morphology and compressive strength of the implant, the supposed modification will be promising for PEKK-based bone implants.

Finally, osteogenic differentiation of the cells incubated with the fabricated samples was investigated. No differentiated cells were observed after the incubation with the control non-modified PEKK sample (Figure 2 (bottom row)). In turn, red round coloring typical for the osteogenically differentiated cells [41] was observed for the cells incubated with the samples with immobilized HAp. The observed effect was intensified with the increasing amount of HAp from the results of elemental composition (Table 2). It is well-known, that both $Ca^{2+}$ ions and inorganic phosphate induce the osteogenic differentiation of MSC [42]. Thus, the proposed method for the immobilization of HAp allows PEKK-based implants to promote the osteogenic differentiation of MSC.

**Conclusions**

A new method for the immobilization of hydroxyapatite (HAp) on the surface of additively printed poly(etherketoneketone) (PEKK) implants was developed. The suggested approach includes single stage based on the treatment of the porous PEKK implants with the suspension of HAp particles in 1,1,1,3,3,3-hexafluoropropan-2-ol (HFP)/water mixture. During the treatment process, PEKK surface swells, which allows the immobilization of HAp particles. The proposed approach allowed to preserve the morphological and mechanical characteristics of the porous PEKK implants prepared using FDM-printing technique. At the same time, the adhesion of stem cells was improved after the immobilization of HAp particles. It is demonstrated that the immobilization of HAp particles from 5 wt. % suspension during only 3 minutes provides the enhanced adhesion of stem cells (the number of adhered cells increased 1.7-times) and their osteogenic differentiation. Thus, the presented approach may be useful for the surface modification of PEKK-based bone implants.

**Acknowledgements**

This study was financially supported by the Russian Science Foundation (project # 24-23-00467).

**Conflict of Interest**

The authors declare no conflict of interest.

**Data Availability Statement**

Data available on request from the authors.